\documentclass[prd,twocolumn,showpacs,amsmath,amssymb,flushbottom]{revtex4-1}

\usepackage{graphicx}
\usepackage{epstopdf}
\usepackage{tabularx}
\usepackage{multirow}
\usepackage[colorlinks,citecolor=blue,linkcolor=blue]{hyperref}
\usepackage[pagewise]{lineno}

\begin{document}

\title{
\boldmath
Measurement of $\mathcal B(\psi(3770)\to\gamma \chi_{c1})$
and search for $\psi(3770)\to\gamma \chi_{c2}$
}

\author{
  \begin{small}
    \begin{center}
      M.~Ablikim$^{1}$, M.~N.~Achasov$^{9,a}$, X.~C.~Ai$^{1}$,
      O.~Albayrak$^{5}$, M.~Albrecht$^{4}$, D.~J.~Ambrose$^{44}$,
      A.~Amoroso$^{48A,48C}$, F.~F.~An$^{1}$, Q.~An$^{45}$,
      J.~Z.~Bai$^{1}$
% ...
      \\
      \vspace{0.2cm}
      (BESIII Collaboration)\\
      \vspace{0.2cm} {\it
        $^{1}$ Institute of High Energy Physics, Beijing 100049, People's Republic of China\\
% ...
$^{4}$ Bochum Ruhr-University, D-44780 Bochum, Germany\\
        $^{5}$ Carnegie Mellon University, Pittsburgh, Pennsylvania 15213, USA\\
% ...
$^{9}$ G.I. Budker Institute of Nuclear Physics SB RAS (BINP), Novosibirsk 630090, Russia\\
% ...
$^{44}$ University of Rochester, Rochester, New York 14627, USA\\
        $^{45}$ University of Science and Technology of China, Hefei 230026, People's Republic of China\\
% ...
$^{48}$ (A)University of Turin, I-10125, Turin, Italy; (B)University of Eastern Piedmont, I-15121, Alessandria, Italy; (C)INFN, I-10125, Turin, Italy\\
% ...
$^{a}$ Also at the Novosibirsk State University, Novosibirsk, 630090, Russia\\
% ...
      }\end{center}
    \vspace{0.4cm}
  \end{small}
}

\affiliation{}

\date{\today}

\begin{abstract}
We report a measurement of the branching fraction for $\psi(3770)\to\gamma\chi_{c1}$ and
search for the transition $\psi(3770)\rightarrow \gamma \chi_{c2}$
based on 2.92~fb$^{-1}$ of $e^+e^-$ data accumulated at
$\sqrt{s}=3.773$~GeV with the BESIII detector at the BEPCII collider.
We measure
$\mathcal B(\psi(3770) \rightarrow \gamma \chi_{c1})=(2.48 \pm 0.15 \pm 0.23) \times 10^{-3}$,
which is the most precise measurement to date.
The upper limit on the branching fraction of
$\psi(3770)\rightarrow \gamma \chi_{c2}$ at a $90\%$ confidence level is
$\mathcal B(\psi(3770) \rightarrow \gamma \chi_{c2})<0.64 \times 10^{-3}$.
The corresponding partial widths are
$\Gamma(\psi(3770) \to \gamma \chi_{c1}) =(67.5\pm 4.1\pm 6.7)$~keV
and $\Gamma(\psi(3770) \to \gamma \chi_{c2}) < 17.4$~keV.
\end{abstract}



\maketitle

\section{Introduction}

The $\psi(3770)$ resonance is the lowest-mass $c\bar c$ state lying
above the open charm-pair threshold (3.73~GeV$/c^2$).
Since its width is 2 orders of magnitude larger than
that of the $\psi(3686)$ resonance, it is traditionally expected to
decay to $D\bar D$ meson pairs
with a branching fraction of more than $99\%$~\cite{Eithtin_chmonuim_prd1978}.
This would be consistent with other conventional mesons lying in the energy region between the open-charm
and open-bottom thresholds.
However, if a meson lying in this region
contains not only a $c\bar c$ pair but also a number of constituent gluons
or additional light quarks and antiquarks,
it may more easily decay to non-$D\bar D$ final states
(such as a lower-mass $c\bar c$ pair plus pions~\cite{Y4260_hybrid}
or light hadrons~\cite{Voloshin_Charmonium})
than conventional mesons.
In addition, if there are some unknown conventional or unconventional mesons
nearby the $c\bar c$ state under study, the measured
non-open-charm-pair decay branching fraction of the $c\bar c$ state
could also be large~\cite{RongG_CPC_34_778_Y2010}.
For this reason, searching for non-open-charm-pair decays of the mesons
lying in this region has become a way to search for unconventional mesons.

In 2003, the BES Collaboration found the first non-open-charm-pair final
state of $J/\psi\pi^+\pi^-$~\cite{hepnp28_325,plb605_63}
in data taken at 3.773~GeV.  Since the final state $J/\psi\pi^+\pi^-$ cannot be directly produced
in $e^+e^-$ annihilation,
this process is interpreted to be a hadronic transition $\psi(3770) \rightarrow J/\psi\pi^+\pi^-$,
although it has not been excluded that this final state may be a decay product
of some other possible structures~\cite{bes2_prl_2structures} which may exist
in this energy region.
Following this observation, the CLEO Collaboration found
that $\psi(3770)$ can also decay into $J/\psi\pi^0\pi^0$, $J/\psi\eta$~\cite{prl96_082004},
$\gamma\chi_{c0}$~\cite{prd74_031106},
$\gamma\chi_{c1}$~\cite{prl96_182002} and $\phi\eta$~\cite{prd73_012002}.
In the CLEO-c measurements, the $\chi_{c0}$ and $\chi_{c1}$ were reconstructed
with $\chi_{c0}\to$ light hadrons and $\chi_{c1}\to\gamma J/\psi$, respectively.
These observations stimulate strong interest in studying other non-$D\bar D$ decays of the $\psi(3770)$,
as well as searching for non-open-charm-pair decays of
other mesons lying in the energy region between the open charm-pair
and open bottom-pair thresholds,
particularly searching for $J/\psi {\rm X}$ or $c\bar c {\rm X}$
(where X denotes any other particle, or  $n\pi$, $n$K, and  $\eta$, where $n=1,2,3\ldots$)
decays of these mesons in this energy region.

Within an $S$-$D$ mixing model, the $\psi(3770)$ resonance
is assumed to be predominantly the $1^3D_1$ $c\bar c$ state with a small admixture of the $2^3S_1$ state.
Based on this assumption, Refs.~\cite{prd44_3562,prd64_094002,prd69_094019,prd72_054026} predict
the partial widths of $\psi(3770)$ $E1$ radiative transitions,
but with large uncertainties. For example,
the partial widths for $\psi(3770)\to \gamma\chi_{c1}$ and
$\psi(3770)\to \gamma\chi_{c2}$ range from 59 to 183~keV
and from 3 to 24~keV, respectively.
In addition,
the transition $\psi(3770)\to \gamma\chi_{c2}$ has yet to be
observed.
Therefore, precision measurements of partial widths of the
$\psi(3770)\to\gamma\chi_{c1,2}$ processes
are critical to test the above mentioned models, and
to better understand the nature of the $\psi(3770)$, as well as to find
the origin of the non-$D\bar D$ decays of the $\psi(3770)$.

In this paper, we report a measurement of
the branching fraction for the transition $\psi(3770) \rightarrow \gamma \chi_{c1}$
and search for the transition $\psi(3770) \rightarrow \gamma \chi_{c2}$
based on $(2916.94\pm29.17)$~pb$^{-1}$ of $e^+e^-$ data~\cite{bes3_lum} taken at $\sqrt{s}=3.773$~GeV
with the BESIII detector~\cite{bes3}
operated at the BEPCII collider.

\section{BESIII Detector}

The BESIII~\cite{bes3} detector is a cylindrical detector with a
solid-angle coverage of $93\%$ of $4\pi$ that operates at the BEPCII~\cite{bes3}
$e^+e^-$ collider.  It consists of several main
components.  A 43-layer main drift chamber (MDC) surrounding the beam
pipe performs precise determinations of charged particle
trajectories and provides ionization energy loss ($dE/dx$)
measurements that are used for charged-particle identification.  An
array of time-of-flight counters (TOF) is located radially outside of
the MDC and provides additional charged particle identification
information.  The time resolution of the TOF system is 80~ps (110~ps) in the
barrel (end-cap) regions,
corresponding to better than 2$\sigma$  $K/\pi$ separation for momenta below about 1 GeV/c.
The solid angle coverage of the barrel TOF is $|\cos \theta|<0.83$,
while that of the end cap is $0.85<|\cos \theta|<0.95$,
where $\theta$ is the polar angle.
A CsI(Tl) electromagnetic calorimeter (EMC) surrounds the
TOF and is used to measure the energies of photons and electrons.
The angular coverage of the barrel EMC is
$|\cos \theta| <0.82$. The two end caps cover
$0.83<|\cos \theta|<0.93$.
A solenoidal superconducting magnet located outside the EMC provides a 1
T magnetic field in the central tracking region of the detector.  The
iron flux return of the magnet is instrumented with about 1200~m$^2$ of
resistive plate muon counters (MUC) arranged in nine layers in the barrel
and eight layers in the end caps that are used to identify muons with
momentum greater than 500~MeV/$c$.

The BESIII detector response is studied using samples of Monte Carlo (MC)
simulated events which are simulated with a {\sc geant4}-based~\cite{geant4} detector simulation software
package, {\sc boost}~\cite{BOOST}.
The production of the $\psi(3770)$ resonance is simulated with the
Monte Carlo event generator $\mathcal{KK}$, {\sc kkmc}~\cite{kkmc}.
The decays of $\psi(3770)\to\gamma\chi_{cJ}$ ($J=0,1,2$) are generated with {\sc EvtGen}~\cite{besevtgen} according to the expected angular
distributions~\cite{angular_model}.
In order to study possible backgrounds,
Monte Carlo samples of inclusive $\psi(3770)$ decays,
$e^+e^-\to(\gamma)J/\psi$, $e^+e^-\to(\gamma)\psi(3686)$,
and $e^+e^-\to q\bar q$ ($q=u,d,s$) are also generated.
For inclusive decays of $\psi(3770)$, $\psi(3686)$ and $J/\psi$,
the known decay modes are generated by {\sc EvtGen} with branching fractions
taken from the PDG~\cite{pdg2014},
while the remaining unknown decay modes are modeled by
{\sc LundCharm}~\cite{lundcharm}.
In addition,
the background process $e^+e^-\to\tau^+\tau^-$ is generated with {\sc kkmc},
while the backgrounds from
$e^+e^-\to(\gamma)e^+e^-$ and $e^+e^-\to(\gamma)\mu^+\mu^-$
are generated with the generator {\sc babayaga}~\cite{babayaga}.

\section{Analysis}

In this analysis, the process $\psi(3770) \rightarrow \gamma \chi_{cJ}$ ($J=1,2$) is reconstructed
using the decay chain $\chi_{cJ} \rightarrow \gamma J/\psi$, $J/\psi\to\ell^+\ell^-$ ($\ell=e$ or
$\mu$).

\subsection{Event selection}

Events that contain two good photon candidates and
exactly two oppositely charged tracks
are selected for further analysis.
For the selection of photons, the deposited energy of a neutral cluster in the EMC
is required to be greater than 50~MeV.
Time information from the EMC is used to
suppress electronic noise and energy deposits unrelated to the event.
To exclude false photons originating from charged tracks,
the angle between the photon candidate and the nearest charged track
is required to be greater than $10^{\circ}$.
Charged tracks are reconstructed from hit patterns in the MDC.
For each charged track, the polar angle $\theta$ is required to satisfy $|\cos\theta|<0.93$.
All charged tracks are required to have a distance of closest approach
to the average $e^+e^-$ interaction point that is less than 1.0~cm
in the plane perpendicular to the beam and less than 15.0~cm along the beam direction.
Electron and muon candidates can be well separated with the
ratio $E/p$,
where $E$ is the energy deposited in the EMC
and $p$ is the momentum measured in the MDC.
If the ratio $E/p$ is greater than 0.7, the charged track is identified as an electron or positron.
Otherwise, if the energy deposited in the EMC is
in the range from 0.05 to 0.35~GeV, the charged track is identified as a muon.
The $J/\psi$ candidates are reconstructed from pairs of leptons with
momenta in a range from 1.2 to 1.9~GeV$/c$.

In the selection of the $\gamma\gamma e^+e^-$ mode, we further require that the cosine of the polar
angle of the positron and electron, $\theta_{e^+}$ and $\theta_{e^-}$, satisfy $\cos\theta_{e^+}<0.5$
and $\cos\theta_{e^-}>-0.5$ to reduce the number of background events from radiative Bhabha scattering.

To exclude background events from $J/\psi\pi^0$ and $J/\psi\eta$
with $\pi^0\to\gamma\gamma$ and $\eta\to\gamma\gamma$,
the invariant mass of the two photons is required to be outside of the $\pi^0$
mass window (0.124, 0.146)~GeV$/c^2$ and the $\eta$ mass window (0.537, 0.558)~GeV$/c^2$.

\subsection{Kinematic fit and mass spectrum of $\gamma J/\psi$}

In order to both reduce background and improve the mass resolution,
a kinematic fit is performed under the $\gamma\gamma\ell^+\ell^-$ hypothesis.
We constrain the total energy and the components of the total momentum
to the expected center-of-mass energy and the three-momentum,
taking into account the small beam crossing angle. In addition to these,
we constrain the invariant mass of the $\ell^+\ell^-$ pair to the $J/\psi$ mass.
If the $\chi^2$ of the 5-constraint (5C) kinematic fit is less than 25,
the event is kept for further analysis.

\begin{figure}[!hbt]
\includegraphics[width=0.5\textwidth]{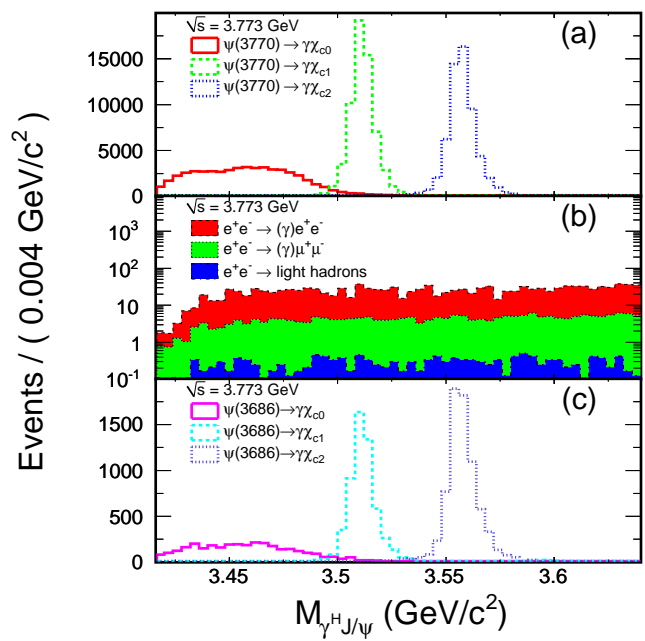}
\caption{
Invariant mass spectra of the selected $\gamma^{H}J/\psi$
combinations from Monte Carlo events generated at $\sqrt s = 3.773$~GeV,
(a) is for the events from $\psi(3770) \rightarrow \gamma \chi_{cJ}\to\gamma\gamma J/\psi\to \gamma\gamma \ell^+\ell^-$ decays,
(b) is for the background events,
and
(c) is the $e^+e^-\to(\gamma_{\rm ISR})\psi(3686)$, $\psi(3686)\rightarrow\gamma \chi_{cJ}\to\gamma\gamma J/\psi\to\gamma\gamma\ell^+\ell^-$
events.
}
\label{mass_ghjpsi_mc}
\end{figure}

The energy of the $\gamma$ from the transition $\psi(3770) \rightarrow \gamma \chi_{cJ}$ for $J=1,2$
is lower than that of the $\gamma$ from the subsequent transition $\chi_{cJ} \rightarrow \gamma J/\psi$,
while the energy of the $\gamma$ from the transition $\psi(3770) \rightarrow \gamma \chi_{c0}$
is usually higher than that of the $\gamma$ from the subsequent transition $\chi_{c0} \rightarrow \gamma J/\psi$.
To reconstruct the $\chi_{c1}$ and $\chi_{c2}$ from the radiative decay of the $\psi(3770)$,
we examine the invariant mass of $\gamma^H J/\psi$,
where $\gamma^H$ refers to the higher energetic photon in the final state $\gamma\gamma\ell^+\ell^-$.
Figure~\ref{mass_ghjpsi_mc} (a) shows the distribution
of the invariant masses of $\gamma^{H} J/\psi$
from the Monte Carlo events of $\psi(3770) \rightarrow \gamma \chi_{cJ}\to\gamma\gamma J/\psi\to\gamma\gamma\ell^+\ell^-$,
which were generated at $\sqrt{s}=3.773$~GeV.
Due to the wrong combination of the photon and $J/\psi$,
the transition $\psi(3770)\to\gamma\chi_{c0}$ produces a broad distribution on the lower side;
the events shown in the peak located at $\sim 3.51$~GeV$/c^2$ are
from the $\psi(3770) \rightarrow \gamma \chi_{c1}$ decay;
while the events from the peak located at $\sim 3.56$~GeV$/c^2$ are
from the $\psi(3770) \rightarrow \gamma \chi_{c2}$ decay.

\begin{figure}[!hbt]
\includegraphics[width=0.5\textwidth]{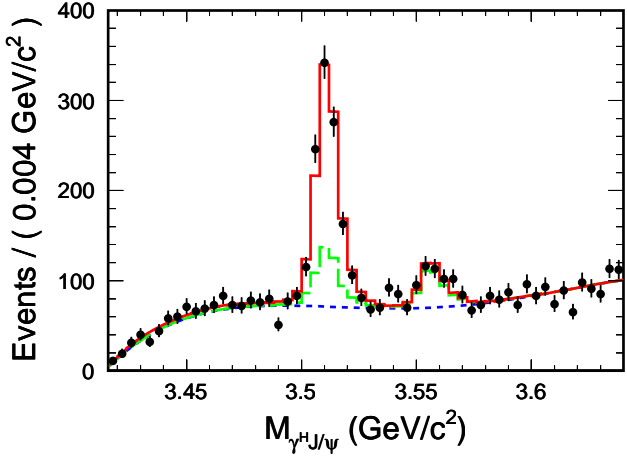}
\caption{
Invariant mass spectrum of the $\gamma^{\rm H}J/\psi$ combinations selected from data.
The dots with error bars represent the data.
The solid (red) line shows the fit.
The dashed (blue) line shows the smooth background.
The long-dashed (green) line is the sum of the smooth background
and the contribution from $e^+e^-\to(\gamma_{\rm ISR})\psi(3686)$ production.
}
\label{mass_ghjpsi_data}
\end{figure}
Figure~\ref{mass_ghjpsi_data} shows the invariant-mass distribution
of $\gamma^{H} J/\psi$ from the data.
There are two clear peaks corresponding to the $\chi_{c1}$ (left) and
the $\chi_{c2}$ (right) signals.
Due to the small branching fraction ($\sim 1\%$) and the wrong combination of the photon and $J/\psi$,
the events from $\chi_{c0}\to\gamma J/\psi$ decays are not clearly observed in Fig.~\ref{mass_ghjpsi_data}.

\subsection{Background studies}

In the selected candidate events,
there are both signal events
for $\psi(3770) \rightarrow \gamma \chi_{cJ}\to\gamma\gamma J/\psi$
and background events.
These background events originate from several sources, including
(1) decays of the $\psi(3770)$ other than the signal modes in question,
(2) $e^+e^-\to(\gamma)e^+e^- $, $e^+e^-\to(\gamma) \mu^+\mu^-$ and $e^+e^-\to(\gamma)\tau^+\tau^-$,
    where the $\gamma$ in parentheses denotes the inclusion of photons from initial state radiation (ISR)
    and final state radiation (FSR),
(3) continuum light hadron production,
(4) ISR $J/\psi$ events,
(5) cross contamination between the $e^+e^-$ and $\mu^+\mu^-$ modes of the signal events, and
(6) $e^+e^-\to(\gamma_{\rm ISR})\psi(3686)$ events produced at $\sqrt{s} = 3.773$~GeV,
    where the notation ``$\gamma_{\rm ISR}$'' denotes the inclusion of produced
    $\psi(3686)$ due to radiative photon in the initial state.

Figure~\ref{mass_ghjpsi_mc} (b) shows different components of the selected $\gamma\gamma J/\psi$ events
misidentified from the Monte Carlo simulated background events for $e^+e^- \to (\gamma) e^+e^-$,
$e^+e^- \to (\gamma) \mu^+\mu^-$,
and
continuum light hadron production,
which are generated at $\sqrt{s}=3.773$~GeV.
The shape of the invariant-mass distribution for these background events
can be well described with a polynomial function.
Using MC simulation,
the contributions from decays of the $\psi(3770)$ other than the signal mode,
$e^+e^- \to (\gamma) \tau^+\tau^-$, ISR $J/\psi$ events,
and cross contamination between the $e^+e^-$ and $\mu^+\mu^-$ modes of the signal events
are found to be negligible.

In addition to the
backgrounds described above, the background events
from $e^+e^-\to(\gamma_{\rm ISR})\psi(3686)$ with $\psi(3686)\to\gamma\chi_{cJ}$ ($\chi_{cJ}\to\gamma J/\psi$,
$J/\psi\to\ell^+\ell^-$) decays can also satisfy the event selection criteria.
This kind of background produced near $\sqrt s = 3.773$ GeV has
the same event topology as that of $\psi(3770)\to\gamma\chi_{cJ}$ decays
and are indistinguishable from the signal events.
The number of background events from $\psi(3686)$ decays can be estimated using
\begin{linenomath*}
\begin{eqnarray}\label{Eq_Nbkg_psip}
  N_{\psi(3686)\to\gamma\chi_{cJ}} &=&  \sigma^{\rm obs}_{\psi(3686)\to\gamma\chi_{cJ}} \times \mathcal L \times \mathcal B_{\chi_{cJ}\to\gamma J/\psi}
  \nonumber \\
  &\times& \mathcal B_{J/\psi\to\ell^+\ell^-} \times \eta_{\psi(3686)\to\gamma\chi_{cJ}},
\end{eqnarray}
\end{linenomath*}
where
$\sigma^{\rm obs}_{\psi(3686)\to\gamma\chi_{cJ}}$ is the
observed cross section of $e^+e^-\to \gamma_{\rm ISR}\psi(3686)$
with $\psi(3686)\to\gamma\chi_{cJ}$ at $\sqrt{s}=3.773$ GeV,
$\mathcal L$ is the integrated luminosity of the data used in the analysis,
$\mathcal B_{\chi_{cJ}\to\gamma J/\psi}$
is the decay branching fraction
of $\chi_{cJ}\to\gamma J/\psi$, $\mathcal B_{J/\psi\to\ell^+\ell^-}$
is
the sum of branching fractions of $J/\psi\to e^+e^-$ and $J/\psi\to\mu^+\mu^-$ decays,
and
$\eta_{\psi(3686)\to\gamma\chi_{cJ}}$ represents the rate of misidentifying
the $\psi(3686)\to\gamma\chi_{cJ}$ events as $\psi(3770)\to\gamma\chi_{cJ}$ signal events.
The observed
cross section for $e^+e^-\to \gamma_{\rm ISR}\psi(3686)\to\gamma\chi_{cJ}$
at $\sqrt{s}$ is obtained with
\begin{linenomath*}
\begin{align}
& \sigma^{\rm obs}_{\psi(3686)\to\gamma\chi_{cJ}} \nonumber \\
& = \int \sigma^{\rm D}_{\psi(3686)\to \gamma\chi_{cJ}}(s^\prime) f(s^\prime) F(x,s) G(s,s^{\prime\prime}) ds^{\prime\prime} dx,
\label{Eq_XSEC_psip_obs}
\end{align}
\end{linenomath*}
where
$\sigma^{\rm D}_{\psi(3686)\to \gamma\chi_{cJ}}(s^\prime)$ is the dressed
cross section for $\psi(3686)\to \gamma\chi_{cJ}$ decay,
$s^\prime=s(1-x)$ is the square of the actual center-of-mass energy of the $e^+e^-$ after radiating the photons,
$x$ is the fraction of the radiative energy to the beam energy,
$f(s^\prime)$ is a phase space factor,
$F(x,s)$ is the sampling function for the radiative energy fraction $x$
at $\sqrt s$~\cite{Structure_Function},
$G(s,s^{\prime\prime})$ is a Gaussian function describing the distribution of the $e^+e^-$ collision
energy with an energy spread $\sigma_E=1.37$ MeV at BEPCII.
$\sigma^{\rm D}_{\psi(3686)\to \gamma\chi_{cJ}}(s^\prime)$
is calculated with
\begin{linenomath*}
\begin{align}
& \sigma^{\rm D}_{\psi(3686)\to \gamma\chi_{cJ}}(s^{\prime}) \nonumber \\
& = \frac{12\pi\Gamma^{ee}_{\psi(3686)}\Gamma^{\rm tot}_{\psi(3686)} \mathcal B(\psi(3686)\to\gamma\chi_{cJ})}
{({s^\prime}^2-M_{\psi(3686)}^2)^2 + (\Gamma^{\rm tot}_{\psi(3686)}M_{\psi(3686)})^2},
\label{Eq_XSEC_psip_B}
\end{align}
\end{linenomath*}
where
$\Gamma^{ee}_{\psi(3686)}$ and $\Gamma^{\rm tot}_{\psi(3686)}$ are, respectively,
the leptonic and total width of the $\psi(3686)$,
$M_{\psi(3686)}$ is the mass of the $\psi(3686)$,
and
$\mathcal B(\psi(3686)\to\gamma\chi_{cJ})$
denotes the decay branching fraction of $\psi(3686)\to\gamma\chi_{cJ}$ ($J=0,1,2$).
The phase space factor is equal to~\cite{fcor}
\begin{linenomath*}
\begin{equation}\label{Eq_fphsp}
    f(s^\prime) = ( {E_\gamma(s^\prime)}/{E_\gamma^0} )^3,
\end{equation}
\end{linenomath*}
where
$E_\gamma(s^\prime)$ and $E_\gamma^0$ are the energies of the photon
in the $\psi(3686)\to\gamma\chi_{cJ}$ decay at $e^+e^-$ energies of $\sqrt{s^\prime}$
and $M_{\psi(3686)}$, respectively.
The rates $\eta_{\psi(3686)\to\gamma\chi_{cJ}}$ of misidentifying $\psi(3686)\to\gamma\chi_{cJ}$
as $\psi(3770)\to\gamma\chi_{cJ}$
are $4.16\times10^{-3}$, $6.88\times10^{-3}$ and $8.86\times10^{-3}$ for
$\chi_{c0}$, $\chi_{c1}$ and $\chi_{c2}$, respectively,
which are estimated with
Monte Carlo simulated events for $\psi(3686)\to\gamma\chi_{cJ}$
generated at $\sqrt{s}=3.773$ GeV.
With
the parameters of the $\psi(3686)$ ($M_{\psi(3686)}=3686.109^{+0.012}_{-0.014}$ MeV,
$\Gamma^{\rm tot}_{\psi(3686)}=299\pm8$ keV and $\Gamma^{ee}_{\psi(3686)}=2.36\pm0.04$ keV),
the luminosity of the data,
the decay branching fractions and the misidentification rates,
we obtain
the numbers of background events from
$\psi(3686)\to\gamma\chi_{cJ}\to\gamma\gamma J/\psi\to\gamma\gamma\ell^+\ell^-$ decays
to be
$5.3   \pm  0.3$ $\chi_{c0}$,
$225.4 \pm 11.7$ $\chi_{c1}$
and
$158.4 \pm  8.5$ $\chi_{c2}$,
where the errors are mainly due to
the uncertainties of the $\psi(3686)$ resonance parameters,
the luminosity,
the branching fractions of $\psi(3686)\to\gamma\chi_{cJ}$, $\chi_{cJ}\to\gamma J/\psi$ and $J/\psi\to\ell^+\ell^-$ decays.

\subsection{Signal events for $\psi(3770)\rightarrow$ $\gamma\chi_{cJ}$}

To extract the number of signal events,
we fit the invariant-mass spectrum
of $\gamma^{H} J/\psi$ shown in Fig.~\ref{mass_ghjpsi_data}
with a function describing the shape of the mass spectrum.
The function is constructed with the Monte Carlo simulated signal shape
as shown in Fig.~\ref{mass_ghjpsi_mc} (a) to describe the signal, a
fourth-order polynomial for the smooth background, and the
Monte Carlo simulated mass shape
for the $e^+e^-\to(\gamma_{\rm ISR})\psi(3686)$ process
with a yield fixed to the predicted size of the corresponding peaking background.
In the fit the expected number of $\psi(3770)\to\gamma\chi_{c0}$ is fixed at $60.1 \pm 8.6$ events,
which is estimated with the branching fraction
for $\psi(3770) \to \gamma \chi_{c0}$ decay~\cite{pdg2014}
and the total number of $\psi(3770)$
as well as the reconstruction efficiency.
The error in the estimated number of events is from the uncertainties of the branching fractions
for $\psi(3770) \to \gamma \chi_{c0}$,
$\chi_{c0}\to\gamma J/\psi$ and $J/\psi\to\ell^+\ell^-$~\cite{pdg2014},
the total number of $\psi(3770)$ and the reconstruction efficiency.

The fit returns $654.2\pm40.3$ and $34.7\pm29.4$ signal events
for $\psi(3770) \to \gamma \chi_{c1}$ and $\psi(3770) \to \gamma \chi_{c2}$ decays, respectively.
The red solid line in Fig.~\ref{mass_ghjpsi_data} shows the best fit.
To estimate the statistical significance of observing
$\psi(3770) \to \gamma \chi_{c2}$ signal events,
we perform a fit with the $\chi_{c2}$ signal amplitude fixed at zero.
Transforming the ratio of the fit likelihoods into the number of
standard deviations at which the null hypothesis can be excluded
gives a statistical signal significance of $1.2$ standard deviations.

\section{Result}

\subsection{Total number of $\psi(3770)$}

The total number of $\psi(3770)$ produced in the data sample is given by
\begin{linenomath*}
\begin{equation}
    N_{\psi(3770)} = \sigma_{\psi(3770)}^{\rm obs}\times\mathcal L,
\end{equation}
\end{linenomath*}
where $\sigma_{\psi(3770)}^{\rm obs}$ is the total cross section for $\psi(3770)$ production
at 3.773~GeV in $e^+e^-$ annihilation, which includes tree-level and both
ISR and vacuum polarization contributions.
The BES-II Collaboration previously measured the cross section
$\sigma^{\rm obs}_{\psi(3770)}(\sqrt{s})|_{{\sqrt{s} = 3.773~{\rm GeV}}} = (7.15 \pm 0.27 \pm 0.27)$ nb~\cite{psipp_cs},
which was obtained by weighting two independent measurements of this cross section~\cite{bes2_3, psipp_cs_2}.
Using this cross section $\sigma^{\rm obs}_{\psi(3770)}(\sqrt{s})|_{{\sqrt{s} = 3.773~{\rm GeV}}}$ and
the luminosity
of the data~\cite{bes3_lum}, we obtain the total number of $\psi(3770)$ produced in the data sample
to be
\begin{linenomath*}
$$N_{\psi(3770)}=(20.86 \pm 1.13)\times 10^6,$$
\end{linenomath*}
where the error is due to the uncertainties of the total cross section for $\psi(3770)$ production
and the luminosity of the data.

\subsection{Branching fraction}

The branching fractions for $\psi(3770) \to \gamma \chi_{c1}$ and $\psi(3770) \to \gamma \chi_{c2}$ decays
are determined with
\begin{linenomath*}
\begin{align}
& \mathcal B({\psi(3770)\to\gamma\chi_{c1,2}}) = \nonumber \\
& \phantom{=} \frac {N_{\psi(3770)\to\gamma\chi_{c1,2}} }
           {N_{\psi(3770)} \mathcal B_{\chi_{c1,2}\to\gamma J/\psi}\mathcal B_{J/\psi \to \ell^+\ell^-} \epsilon_{\psi(3770) \to \gamma \chi_{c1,2}} },
\label{Eq_Bf}
\end{align}
\end{linenomath*}
where
$N_{{\psi(3770)\to\gamma\chi_{c1,2}}}$ is the observed number of signal events for
$\psi(3770) \to \gamma \chi_{c1,2}$ decays,
$\mathcal B_{\chi_{c1,2}\to\gamma J/\psi}$ is the branching fraction for $\chi_{c1,2}\to\gamma J/\psi$,
$\mathcal B_{J/\psi \to \ell^+\ell^-}$ is the branching fraction for $J/\psi \to \ell^+\ell^-$ decay,
and $\epsilon_{\psi(3770) \to \gamma \chi_{c1,2}}$ is the efficiency
for reconstructing this decay.

The reconstruction efficiencies for observing
$\psi(3770)\to\gamma\chi_{c1}$ and $\psi(3770)\to\gamma\chi_{c2}$ decays
are determined with Monte Carlo simulated events for these decays.
With large Monte Carlo samples, the efficiencies are found to be
$\epsilon_{\psi(3770) \to \gamma \chi_{c1}}=(31.25\pm 0.10)\%$ and
$\epsilon_{\psi(3770) \to \gamma \chi_{c2}}=(28.77\pm 0.10)\%$,
where the errors are statistical.

Inserting the corresponding numbers into Eq.~(\ref{Eq_Bf})
yields the branching fractions
\begin{linenomath*}
\begin{equation}
    \mathcal B(\psi(3770) \to \gamma \chi_{c1}) = (2.48\pm0.15\pm0.23)\times10^{-3},
\end{equation}
\end{linenomath*}
and
\begin{linenomath*}
\begin{equation}\label{bf_gXc2}
    \mathcal B(\psi(3770) \to \gamma \chi_{c2}) = (0.25\pm 0.21\pm 0.18)\times10^{-3},
\end{equation}
\end{linenomath*}
where the first errors are statistical and the second systematic.

The systematic uncertainty in the measured branching fractions of $\psi(3770)\to\gamma\chi_{c1}$ and $\psi(3770)\to\gamma\chi_{c2}$
includes eight contributions:
(1) the uncertainty in the total number of  $\psi(3770)$ ($5.4\%$),
which contains the uncertainty in the observed cross section for $\psi(3770)$
production at $\sqrt{s}=3.773$~GeV~\cite{psipp_cs} and
the uncertainty in the luminosity measurement~\cite{bes3_lum},
(2) the uncertainty in the particle identification ($0.1\%$)
determined by comparing the lepton identification efficiencies for data and
Monte Carlo events, which are measured using
the lepton samples selected from the
$\psi(3686)\to\pi^+\pi^-J/\psi$, $J/\psi\to\ell^+\ell^-$ process,
(3) the uncertainty in the extra $\cos\theta_{e^\pm}$ requirement ($0.1\%$)
estimated by comparing
the acceptances of this requirement for data and Monte Carlo events,
which are determined using the electron samples selected from the $\psi(3686)\to\pi^+\pi^-J/\psi$,
$J/\psi\to e^+e^-$ process,
(4) the uncertainty due to photon selection
($1.0\%$ per photon~\cite{photon}),
(5) the uncertainty associated with the kinematic fit ($2.1\%$)
determined
by comparing the $\chi^2$ distributions and the efficiencies of the $\chi^2<25$ requirement
for data and Monte Carlo simulation,
which are obtained using the $\psi(3686)\to\gamma\gamma\ell^+\ell^-$ events
selected from data taken at $\sqrt{s}=3.686$~GeV and
the corresponding Monte Carlo samples,
(6) the uncertainty in the reconstruction efficiency ($0.3\%$)
arising from the Monte Carlo statistics,
(7) the uncertainties in the branching fractions
of $\chi_{c1,2}\to\gamma J/\psi$ and $J/\psi\to\ell^+\ell^-$ decays
($3.6\%$ for $\gamma \chi_{c1}$, $3.7\%$ for $\gamma \chi_{c2}$~\cite{pdg2014}),
and
(8) the uncertainty associated with the fit to the mass spectrum
($6.1\%$ for $\gamma \chi_{c1}$, $73.2\%$ for $\gamma \chi_{c2}$)
determined by changing the fitting range,
changing the order of the polynomial,
varying the magnitude of the peaking background
from the radiative $\psi(3686)$ tail by $\pm1\sigma$
and
using an alternative signal function (Monte Carlo shape convoluted with a Gaussian function).
These systematic uncertainties are summarized in Table~\ref{Tab:sys_err}.
Adding these systematic uncertainties in quadrature yields total systematic uncertainties
of $9.4\%$ and $73.6\%$ for $\psi(3770) \to \gamma \chi_{c1}$ and
$\psi(3770) \to \gamma \chi_{c2}$ decays, respectively.

\begin{table}[!hbp]
\caption{
Summary of the systematic uncertainties (\%) in the measurements of the
branching fractions for $\psi(3770) \to \gamma \chi_{c1}$ and $\gamma \chi_{c2}$.
}
\label{Tab:sys_err}
\begin{ruledtabular}
\begin{tabular}{lcc}
Source& $\gamma\chi_{c1}$ & $\gamma\chi_{c2}$ \\
\hline
Total number of $\psi(3770)$           & 5.4         &  5.4       \\
Particle identification                & 0.1         &  0.1       \\
$\cos\theta_{e^\pm}$ cut               & 0.1         &  0.1       \\
Photon selection                       & 2.0         &  2.0       \\
Kinematic fit                          & 2.1         &  2.1       \\
Efficiency                             & 0.3         &  0.3       \\
Branching fractions                    & 3.6         &  3.7       \\
Fit to the mass spectrum               & 6.1         & 73.2       \\
\hline
Total                                  & 9.4         & 73.6      \\
\end{tabular}
\end{ruledtabular}
\end{table}

To obtain an upper limit on $\mathcal B(\psi(3770) \to \gamma \chi_{c2})$,
we integrate a likelihood function from zero to the value of
$\mathcal B(\psi(3770) \to \gamma \chi_{c2})$ corresponding to $90\%$ of
the integral from zero to infinity.
The likelihood function is a Gaussian function constructed with
the mean value of $\mathcal B$ and a standard deviation
which includes both the statistical and systematic errors.
Using this method, an upper limit on the branching fraction of $\psi(3770)\to\gamma\chi_{c2}$ is set to
\begin{linenomath*}
\begin{equation}
    \mathcal B(\psi(3770) \to \gamma \chi_{c2}) < 0.64\times10^{-3}
\end{equation}
\end{linenomath*}
at the $90\%$ confidence level (C.L.).

\subsection{Partial width}

Using the total width $\Gamma^{\rm tot}_{\psi(3770)}=(27.2 \pm 1.0)$ MeV~\cite{pdg2014},
we transform the measured branching fractions to the transition widths.
This yields
\begin{linenomath*}
$$\Gamma(\psi(3770)\rightarrow\gamma\chi_{c1})=(67.5 \pm 4.1 \pm 6.7)~{\rm keV}$$
and the upper limit at the $90\%$ C.L.
$$\Gamma(\psi(3770)\rightarrow\gamma\chi_{c2})<17.4~{\rm keV}.$$
\end{linenomath*}
The measured partial widths for these two transitions are
compared to several theoretical predictions in Table~\ref{tab:Cmp_G}.
\begin{table}
\caption{Comparison of measured partial widths with theoretical predictions,
where $\phi$ is the mixing angle of the $S$-$D$ mixing model.}
\label{tab:Cmp_G}
\begin{ruledtabular}
\begin{tabular}{lcc}
\multirow{2}{*}{Experiment/theory} & \multicolumn{2}{c}{$\Gamma(\psi(3770) \to \gamma\chi_{cJ})$ (keV)} \\
 & $J=1$ & $J=2$ \\
\hline
 This work    & $67.5 \pm 4.1 \pm 6.7$        &  $<17.4$ \\
\hline
 Ding-Qin-Chao~\cite{prd44_3562} \\
 ~~ nonrelativistic & 95 & 3.6  \\
 ~~ relativistic     & 72 & 3.0  \\
\hline
 Rosner $S$-$D$ mixing~\cite{prd64_094002}\\
 ~~ $\phi = 12^{\circ}$~\cite{prd64_094002}                          & $73\pm 9$ & $24\pm 4$ \\
 ~~ $\phi =(10.6\pm1.3)^{\circ}$~\cite{BESIII_physics}               & $79\pm 6$ & $21\pm 3$ \\
 ~~ $\phi = 0^{\circ}$ (pure $1^3D_1$ state)~\cite{BESIII_physics}   & 133       & 4.8 \\
\hline
 Eichten-Lane-Quigg~\cite{prd69_094019} \\
 ~~ nonrelativistic               & 183       & 3.2  \\
 ~~ with coupled-channel corr.     & 59        & 3.9  \\
\hline
 Barnes-Godfrey-Swanson~\cite{prd72_054026} \\
 ~~ nonrelativistic                 & 125       & 4.9  \\
 ~~ relativistic                     & 77        & 3.3  \\
\end{tabular}
\end{ruledtabular}
\end{table}

\subsection{Partial cross section}
Using the cross section
$\sigma_{\psi(3770)}=(9.93\pm0.77)$~nb
for $\psi(3770)$ production at $\sqrt{s}=3.773$~GeV,
which is calculated using $\psi(3770)$ resonance parameters~\cite{pdg2014},
together with the measured branching fractions for these two decays,
we obtain the partial cross section for the $\psi(3770)\to\gamma\chi_{c1}$ transition to be
\begin{linenomath*}
$$\sigma(\psi(3770) \to \gamma \chi_{c1}) =(24.6\pm 1.5 \pm3.0)~{\rm pb}$$
and the upper limit at the $90\%$ C.L. on the partial cross section for the $\psi(3770)\to\gamma\chi_{c2}$ transition to be
$$\sigma(\psi(3770) \to \gamma \chi_{c2})<6.4~{\rm pb}.$$
\end{linenomath*}

\section{Summary}

By analyzing 2.92~fb$^{-1}$ of data collected at $\sqrt{s}=3.773$~GeV with
the BESIII detector operated at the BEPCII, we measure
$\mathcal B(\psi(3770) \to \gamma \chi_{c1})=(2.48 \pm 0.15 \pm 0.23) \times 10^{-3}$
and set a $90\%$ C.L. upper limit $\mathcal B(\psi(3770) \to \gamma \chi_{c2}) < 0.64 \times 10^{-3}$.
This measured branching fraction for $\psi(3770) \to \gamma \chi_{c1}$
is consistent within error with
$\mathcal B(\psi(3770) \to \gamma \chi_{c1})=(2.8 \pm 0.5\pm 0.4) \times 10^{-3}$
measured by CLEO-c~\cite{prl96_182002},
but the precision of this measurement is improved by more than a factor of 2.

\begin{acknowledgments}
The BESIII Collaboration thanks the staff of BEPCII and the IHEP computing center for their strong support. This work is supported in part by National Key Basic Research Program of China under Contracts No. 2009CB825204, and No. 2015CB856700; the National Natural Science Foundation of China (NSFC) under Contracts No. 10935007, No. 11125525, No. 11235011, No. 11322544, No. 11335008, and No. 11425524; the Chinese Academy of Sciences (CAS) Large-Scale Scientific Facility Program; the Joint Large-Scale Scientific Facility Funds of the NSFC and CAS under Contracts No. 11179007, No. U1232201, and No. U1332201; CAS under Contracts No. KJCX2-YW-N29, and No. KJCX2-YW-N45; the 100 Talents Program of CAS; INPAC and the Shanghai Key Laboratory for Particle Physics and Cosmology; the German Research Foundation DFG under Contract No. Collaborative Research Center CRC-1044; the Istituto Nazionale di Fisica Nucleare, Italy; the Ministry of Development of Turkey under Contract No. DPT2006K-120470; the Russian Foundation for Basic Research under Contract No. 14-07-91152; the U. S. Department of Energy under Contracts No. DE-FG02-04ER41291, No. DE-FG02-05ER41374, No. DE-FG02-94ER40823, and No. DESC0010118; the U.S. National Science Foundation; the University of Groningen (RuG) and the Helmholtzzentrum fuer Schwerionenforschung GmbH (GSI), Darmstadt; and the WCU Program of the National Research Foundation of Korea under Contract No. R32-2008-000-10155-0.
\end{acknowledgments}


\begin{thebibliography}{99}

\bibitem{Eithtin_chmonuim_prd1978}
E. Eichten, K. Gottfried, T. Kinoshita, K. D. Lane, and T.-M. Yan, Phys. Rev. D {\bf 17}, 3090 (1978).

\bibitem{Y4260_hybrid}
E. Kou and O. Penea, Phys. Lett. B {\bf 631}, 164 (2005).

\bibitem{Voloshin_Charmonium}
M. B. Voloshin, Prog. Part. Nucl. Phys. {\bf 61}, 455 (2008).

\bibitem{RongG_CPC_34_778_Y2010}
G. Rong, Chin. Phys C {\bf 34}, 788 (2010).

\bibitem{hepnp28_325}
J. Z. Bai {\it et al.} (BES Collaboration), Chin. Phys. C {\bf 28}, 325 (2004).

\bibitem{plb605_63}
J. Z. Bai {\it et al.} (BES Collaboration), Phys. Lett. B {\bf 605}, 63 (2005).

\bibitem{bes2_prl_2structures}
M. Ablikim {\it et al.} (BES Collaboration), Phys. Rev. Lett. {\bf 101}, 102004 (2008).

\bibitem{prl96_082004}
N. E. Adam {\it et al.} (CLEO Collaboration), Phys. Rev. Lett. {\bf 96}, 082004 (2006).

\bibitem{prd74_031106}
R. A. Briere {\it et al.} (CLEO Collaboration), Phys. Rev. D {\bf 74}, 031106(R) (2006) .

\bibitem{prl96_182002}
T. E. Coan {\it et al.} (CLEO Collaboration), Phys. Rev. Lett. {\bf 96}, 182002 (2006).

\bibitem{prd73_012002}
G. S. Adams {\it et al.} (CLEO Collaboration), Phys. Rev. D {\bf 73}, 012002 (2006).

\bibitem{prd44_3562}
Y.-B. Ding, D.-H. Qin and K.-T. Chao, Phys. Rev. D {\bf 44}, 3562 (1991).

\bibitem{prd64_094002}
J. L. Rosner, Phys. Rev. D {\bf 64}, 094002 (2001).

\bibitem{prd69_094019}
E. J. Eichten, K. Lane and C. Quigg, Phys. Rev. D {\bf 69}, 094019 (2004).

\bibitem{prd72_054026}
T. Barnes, S. Godfrey and E. S. Swanson, Phys. Rev. D {\bf 72}, 054026 (2005).

\bibitem{bes3_lum}
M. Ablikim {\it et al.} (BESIII Collaboration), Chin. Phys. C {\bf 37}, 123001 (2013).

\bibitem{bes3}
M. Ablikim {\it et al.} (BESIII Collaboration), Nucl. Instrum. Methods Phys. Res. Sect. A {\bf 614}, 345 (2010).

\bibitem{geant4}
S. Agostinelli {\it et al.} (GEANT4 Collaboration), Nucl. Instrum.
Methods Phys. Res., Sect. A {\bf 506}, 250 (2003).

\bibitem{BOOST}
Z. Y. Deng {\it et al.}, Chin. Phys. C {\bf 30}, 371 (2006).

\bibitem{kkmc}
S. Jadach, B. F. L. Ward, and Z. Was, Comput. Phys. Commun. {\bf 130}, 260 (2000).

\bibitem{besevtgen}
D. J. Lange, Nucl. Instrum. Meththods Phys. Res. Sect. A {\bf 462}, 152 (2001);
R.-G. Ping, Chin. Phys. C {\bf 32}, 599 (2008).

\bibitem{angular_model}
E. Eichten, K. Gottfried, T. Kinoshita, J. Kogut, K. D. Lane,
and T.-M. Yan, Phys. Rev. Lett {\bf 34}, 369 (1975);
G. Li, Y. S. Zhu and Z. Y. Wang, Chin. Phys. C {\bf 30}, 718 (2006).

\bibitem{pdg2014}
K. A. Olive {\it et al.} (Particle Data Group), Chin. Phys. C {\bf 38}, 090001 (2014).

\bibitem{lundcharm}
J. C. Chen, G. S. Huang, X. R. Qi, D. H. Zhang and Y. S. Zhu, Phys. Rev. D {\bf 62}, 034003 (2000).

\bibitem{babayaga}
G. Balossini, C. M. Carloni Calame, G. Montagna, O. Nicrosini and F. Piccinini, Nucl. Phys. B {\bf 758}, 227 (2006).

\bibitem{Structure_Function}
E. A. Kuraev and V. S. Fadin, Yad. Fiz. {\bf 41}, 733 (1985); Sov. J. Nucl. Phys. {\bf 41}, 466 (1985).

\bibitem{fcor}
See Eq. (93) in N. Brambilla {\it et al}, Eur. Phys. J. C {\bf 71}, 1534 (2011).

\bibitem{psipp_cs}
M. Ablikim {\it et al.} (BES Collaboration), Phys. Lett. B {\bf 650}, 111 (2007).

\bibitem{bes2_3}
M. Ablikim {\it et al.} (BES Collaboration), Phys. Rev. Lett. {\bf 97}, 121801 (2006).

\bibitem{psipp_cs_2}
M. Ablikim {\it et al.} (BES Collaboration), Phys. Lett. B {\bf 652}, 238 (2007).

\bibitem{photon}
M. Ablikim {\it et al.} (BESIII Collaboration), Phys. Rev. D {\bf 81}, 052005 (2010).

\bibitem{BESIII_physics}
G. Rong, D. H. Zhang, and H. L. Ma, ``$\psi(3770)$ non-$D\bar{D}$ Decays'', in
{\it Physics at BES-III}, edited by Kuang-Ta Chao and Yifang Wang, Int. J. Mod. Phys. A {\bf 24} Supplement 1 (2009), p. 414.


\end{thebibliography}
\end{document}